\documentclass[pdflatex,sn-nature]{sn-jnl}

\usepackage{graphicx}%
\usepackage{multirow}%
\usepackage{amsmath,amssymb,amsfonts}%
\usepackage{amsthm}%
\usepackage{mathrsfs}%
\usepackage[title]{appendix}%
\usepackage{xcolor}%
\usepackage{textcomp}%
\usepackage{manyfoot}%
\usepackage{booktabs}%
\usepackage{algorithm}%
\usepackage{algorithmicx}%
\usepackage{algpseudocode}%
\usepackage{listings}%
\usepackage[T1]{fontenc}
\usepackage{float}
\usepackage[utf8]{inputenc}
\usepackage{caption}
\usepackage{lineno}

\newcounter{extendedfigure}
\renewcommand{\theextendedfigure}{\arabic{extendedfigure}}

\newcommand{\extendedfigcaption}[2]{%
    \refstepcounter{extendedfigure}%
    \caption*{%
        \textbf{Extended Data Fig. \theextendedfigure\ | #1.} #2%
    }%
}

\theoremstyle{thmstyleone}%
\theoremstyle{thmstyletwo}%

\theoremstyle{thmstylethree}%

\unnumbered

\begin{document}

\title{Disentangling spin polarization from driven circular ionic motion in EuTiO$_3$}


\author*[1]{\fnm{Clifford J.} \sur{Allington}}\email{clifford.allington@psi.ch}

\author[2]{\fnm{Matthew J.} \sur{Lutz}}

\author[2]{\fnm{Enoch (Sin-Hang)} \sur{Ho}}

\author[3]{\fnm{Fabian} \sur{Graf}}

\author[3]{\fnm{Martina} \sur{Basini}}

\author[1]{\fnm{Hiroki} \sur{Ueda}}

\author[1]{\fnm{Michael} \sur{Grimes}}

\author[2]{\fnm{Alexander B.} \sur{Elliott}}

\author[2]{\fnm{Megan F.} \sur{Biggs}}

\author[2]{\fnm{Ravi} \sur{Finn}}

\author[1]{\fnm{Shih-Wen} \sur{Huang}}

\author[4]{\fnm{Merideth A.} \sur{Henstridge}}

\author[4]{\fnm{Matthias C.} \sur{Hoffmann}}

\author[4]{\fnm{Takahiro} \sur{Sato}}

\author[4]{\fnm{Roberto} \sur{Alonso-Mori}}

\author[4]{\fnm{Diling} \sur{Zhu}}

\author[4]{\fnm{Quynh L.} \sur{Nguyen}}

\author[4]{\fnm{Vincent} \sur{Esposito}}

\author[4]{\fnm{Jeffrey T.} \sur{Babicz Jr.}}

\author[1]{\fnm{Elizabeth} \sur{Skoropata}}

\author[1]{\fnm{Biaolong} \sur{Liu}}

\author[1]{\fnm{Eugenio} \sur{Paris}}

\author[1]{\fnm{Arnau Romaguera} \sur{Camps}}

\author[1]{\fnm{Elia} \sur{Razzoli}}

\author[1]{\fnm{Roman} \sur{Mankowsky}}

\author[5]{\fnm{Flavio} \sur{Capotondi}}

\author[6]{\fnm{Nicolas} \sur{Jaouen}}

\author[7,8]{\fnm{Daniele} \sur{Pergolesi}}

\author[1]{\fnm{Milan} \sur{Radovic}}

\author[1,3]{\fnm{Matteo} \sur{Savoini}}

\author[1,3]{\fnm{Steven L.} \sur{Johnson}}

\author[2]{\fnm{Jeremy A.} \sur{Johnson}}

\author*[1]{\fnm{Urs} \sur{Staub}}\email{urs.staub@psi.ch}

\affil*[1]{\orgdiv{Center for Photon Science}, \orgname{Paul Scherrer Institut}, \orgaddress{\street{Forschungsstrasse 111}, \city{Villigen}, \postcode{5232}, \country{Switzerland}}}

\affil[2]{\orgdiv{Department of Chemistry and Biochemistry}, \orgname{Brigham Young University}, \orgaddress{\city{Provo}, \postcode{84602}, \state{Utah}, \country{United States of America}}}

\affil[3]{\orgdiv{Department of Physics}, \orgname{ETH Zurich}, \orgaddress{\street{Raemistrasse 101}, \city{Zurich}, \postcode{8092}, \country{Switzerland}}}

\affil[4]{\orgdiv{SLAC National Accelerator Laboratory}, \orgname{Stanford University}, \orgaddress{\street{2575 Sand Hill Rd.}, \city{Menlo Park}, \postcode{94025}, \country{United States of America}}}

\affil[5]{\orgname{Elettra Sincrotrone Trieste}, \orgaddress{\street{Area Science Park T1-T2}, \city{Basovizza TS}, \postcode{34149}, \country{Italy}}}

\affil[6]{ \orgname{Synchrotron SOLEIL}, \orgaddress{\street{L'Orme des Merisiers Departementale 128}, \city{Saint-Aubin}, \postcode{91190}, \country{France}}}

\affil[7]{\orgdiv{Center for Neutrons and Muons Science}, \orgname{Paul Scherrer Institut}, \orgaddress{\street{Forschungsstrasse 111}, \city{Villigen}, \postcode{5232}, \country{Switzerland}}}

\affil[8]{\orgdiv{Center for Energy and Environmental Sciences}, \orgname{Paul Scherrer Institut}, \orgaddress{\street{Forschungsstrasse 111}, \city{Villigen}, \postcode{5232}, \country{Switzerland}}}


\abstract{Coherently driven circular ionic motion has been reported to produce large helicity-dependent optical responses attributed to transient magnetization. However, the microscopic nature and magnitude of this phenomenon, sometimes referred to as dynamical multiferroicity, remain strongly debated. Here, we directly study the connection between ionic circulation induced by a high-field circular terahertz (THz) drive and its effect on the spin system of EuTiO$_3$ using X-rays. This material exhibits an optical response consistent with the putative magnetic signal observed in related non-magnetic materials. Furthermore, the presence of Eu$^{2+}$ ions enables the use of X-ray magnetic circular dichroism (XMCD) to test for the creation of transient spin polarization. Simultaneously, ultrafast X-ray diffraction (XRD) measures the circular ionic motion from which we determine the mechanical angular momentum. We find a classical ionic contribution of $3\times10^{-8}~\mu_B$ from XRD and upper limits of $0.03~\mu_B$ and $0.11~\mu_B$ from the sensitivity of XMCD at the europium $M_5$ and $L_2$ edges, probing the $4f$ and $5d$ shells of Eu$^{2+}$, respectively. These observations show that large-amplitude circular ionic motion in this system is not accompanied by a detectable spin polarization despite the clear signature in the optical data, with the XMCD upper bounds and classical contribution differing by many orders of magnitude.}






\maketitle

\section{Main}\label{sec1}

 In the last decade, phonon modes that carry angular momentum, often referred to as $chiral~phonons$, have emerged as an intriguing platform with which to study the fundamental coupling pathways between the lattice and spin degrees of freedom. Axial phonons, referring to modes with circular motion that lack a significant parallel linear momentum \cite{juraschek2025chiral}, have been of particular interest to investigate these coupling pathways, as they can be directly accessed with recently realized high-field circularly polarized THz light pulses. The coherent excitation of these modes was predicted to produce a transient magnetization due to the angular momentum of charged ions, which was given the name $dynamical~multiferroicity$ in analogy with the mechanism of type II multiferroics \cite{DM2017dynamical}. Following this, experiments were performed that reported very large effective fields at ultrafast timescales when driving axial phonons with high-field circularly polarized terahertz (THz) pulses \cite{basini2024terahertz,hanyu2023large}. These experiments are performed with optical probes, which are sensitive to time-reversal symmetry breaking processes, but are unable to unambiguously determine whether a true non-equilibrium magnetization is induced. This has resulted in a significant discussion regarding the microscopic origin of such signals \cite{merlin2024unraveling, sellati2026light}. In this work, we use X-rays to directly quantify the circular ionic motion and its angular momentum as well as the element-specific magnetic response of Eu$^{2+}$ ions in the material, EuTiO$_3$. This provides quantitative measures of the magnitude of the magnetic moment from the circular ionic motion in EuTiO$_3$, demonstrating that the spin system may be negligibly affected by such processes despite a relatively strong optical response associated with time-reversal symmetry breaking.

 When dynamical multiferroicity was initially proposed, the classical Maxwellian field produced by circular ionic motions was predicted to be a small quantity, with associated moments on the order of a nuclear magneton \cite{DM2017dynamical,juraschek2019orbital}. However, a phenomenological model developed within that framework stated that the gyromagnetic tensor, found from the splitting of degenerate phonon eigenfrequencies in a finite magnetic field via the phonon Zeeman effect, should also govern the size of the moment produced by circular ionic motion. Thus, other microscopic processes could exist to produce large effective fields that are non-Maxwellian in origin, despite the tiny classical contribution from the ionic motion \cite{merlin2024unraveling}. In rare-earth trihalide systems, where a large phonon Zeeman effect and an ultrafast Faraday-like response are observed \cite{hanyu2023large,pz11976observation,pz1981magnetic,pz21977magnetic}, a microscopic origin has been established. The so-called effective field in these systems originates from crystal-field transitions that can directly flip spins during circular THz excitation \cite{chaudhary2024giant,juraschek2022giant}. In contrast, the large magnetic moment of $0.1~\mu_B$ reported in SrTiO$_3$ has generated significant discussion \cite{basini2024terahertz}, as a mechanism for such a large effective field is still debated \cite{merlin2024unraveling, sellati2026light,kto2023electron,nuclearquantum2025ultrafast,phononinverse2024phonon}. 
 
 Here, we investigate the related compound, EuTiO$_3$, as a model system to measure the magnitude of the magnetic moment and spin-lattice coupling induced by circular THz excitation. In bulk, EuTiO$_3$ exhibits quantum paraelectric soft-mode lattice dynamics similar to SrTiO$_3$ \cite{narayan2019multiferroic,kamba2007magnetodielectric,das2012quantum}, but in place of the zero-spin Sr ion it contains Eu$^{2+}$ with spin $\mathbf{S}=7/2$. The large Eu$^{2+}$ spin therefore provides a sensitive element-specific probe of the magnetization through time-resolved X-ray magnetic circular dichroism (XMCD) in the paramagnetic state. The cubic unit cell of bulk EuTiO$_3$ is shown in Fig. 1a and 1b, where the proposed idea of polarizing the spin with circular ionic motion is illustrated. We performed a series of measurements on an epitaxially grown film of EuTiO$_3$ driven by high-field circularly polarized THz pump pulses that drive circular ionic motion (see Methods). The probes for these experiments include ultrafast optical Faraday rotation, X-ray diffraction (XRD) and XMCD. We note that the thin film is ideal from a technical standpoint, since it eliminates penetration depth mismatch of the pump and probe, but that epitaxial strain produces a ferroelectric transition at approximately 250 K and a ferromagnetic phase at temperatures below $\sim$4 K \cite{thinfilm2010strong}. Although experiments reporting dynamical multiferroic signals in other systems have primarily relied on optical probes, time-resolved XRD allows us to directly measure ultrafast structural dynamics \cite{johnson2008nanoscale,beaud2014time,mankowsky2014nonlinear,kozina2019terahertz,johnson2025perspective} and time-resolved XMCD directly probes ultrafast changes in magnetization  \cite{radu2011transient,boeglin2010distinguishing,graves2013nanoscale}. Our combination of these techniques provides quantitative measures of the ionic motion together with XMCD upper bounds on the circular-phonon-driven magnetism in EuTiO$_3$.

\begin{figure}
\centering
\includegraphics[width=1\linewidth]{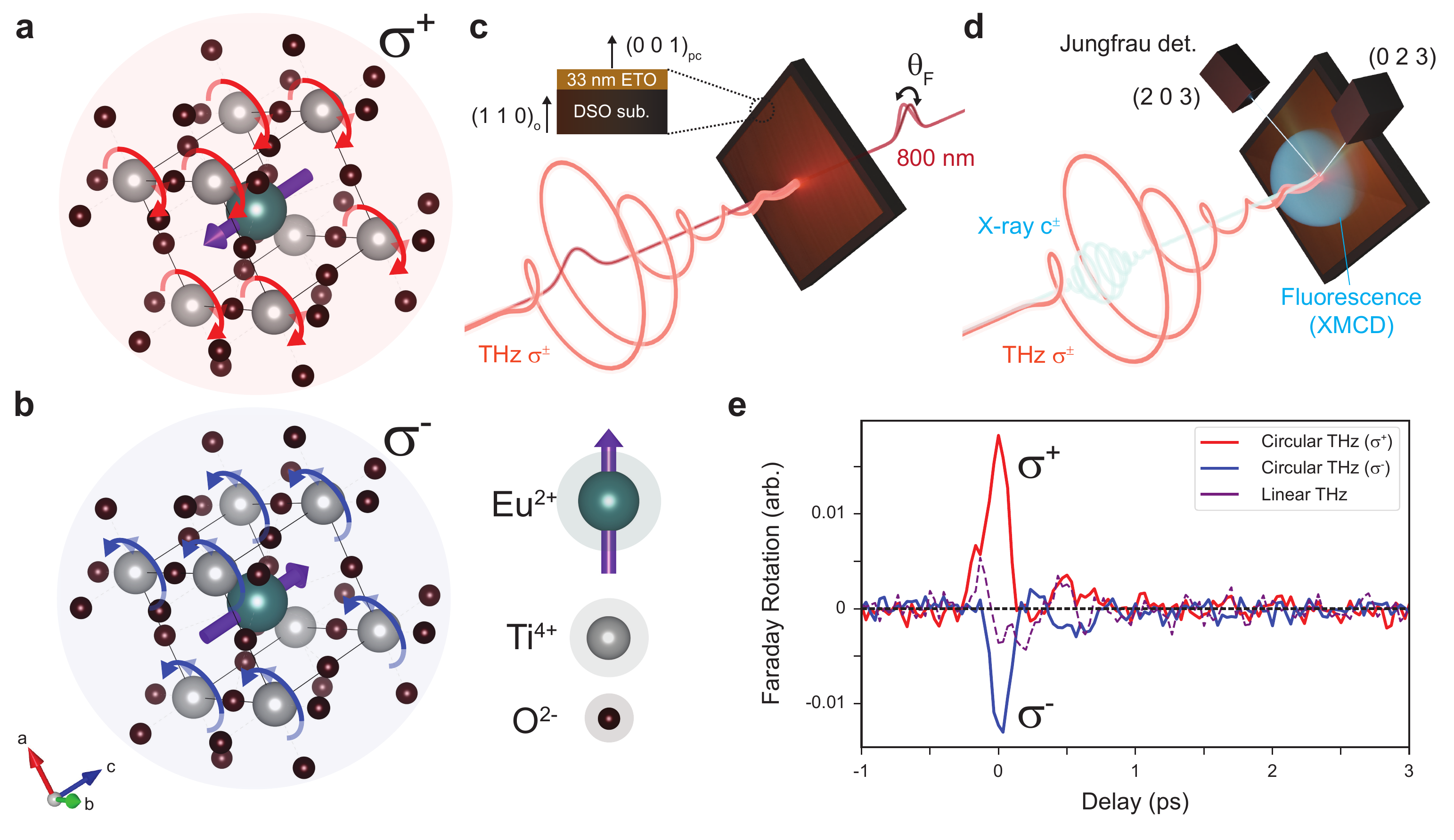}
\caption{\textbf{Optical and X-ray probes of circularly driven EuTiO$_3$.}
(a),(b) Conceptual image of dynamical multiferroicity in the bulk cubic EuTiO$_3$ perovskite cell. Ions circulating clockwise ($\sigma^+$) or counterclockwise ($\sigma^-$) (illustrated for the Ti ions) are proposed to polarize the europium spin in opposite directions. (c) Illustration of 800 nm optical probe polarization (Faraday) rotation with high-field  circular THz pump. The schematic of the epitaxial EuTiO$_3$ (ETO) film on the DyScO$_3$ (DSO) substrate and their out-of-plane orientations. The materials are represented in the pseudocubic (pc) basis for ETO and the orthorhombic basis for DSO. (d) Illustration of X-ray (circularly polarized) diffraction and fluorescence probes with high-field circular THz pump. (e) Faraday rotation signal of the EuTiO$_3$ film during high-field THz excitation of variable polarization. A clear helicity-dependent response is observed when exciting with right- or left-handed THz pulses, in contrast to a linearly polarized excitation.}
\end{figure}

Optical measurements were performed at room temperature with a circular THz-pump pulse and an optical-probe pulse to provide a direct comparison with similar experiments reporting dynamical multiferroicity on other systems. In these experiments, we measure the rotation of the polarization state of a linearly polarized 800 nm probe pulse (see Fig. 1c). This provides a measure of time-reversal symmetry breaking typically attributed to sample magnetization via the Faraday effect. In our experiment, the high-field circular THz pump is generated by combining linear horizontal and vertical pulses with a variable delay, allowing for a controllable elliptical polarization state (see Methods). This approach allows for the isolation of effects that stem from circular polarization as opposed to effects that stem from linear components such as THz-induced birefringence  \cite{biggs2026ultrafast}. In Fig. 1e, the polarization-dependent Faraday response is shown for two helicities of THz-pump, in addition to a linear THz excitation. A clear helicity-dependent signal is observed, which is considerably suppressed for the linear case. The duration of the effect is comparable with the length of the circulating THz pulse (see Fig. 4a). Since the film is only 33 nm thick and the response is not present for the bare substrate (see Supplemental), this demonstrates that the material generates a clear helicity-dependent optical rotation, commonly associated with the phonon-driven magnetism of dynamical multiferroicity.

\section{Mechanical angular momentum}\label{sec2}

\begin{figure}
\centering
\includegraphics[width=1\linewidth]{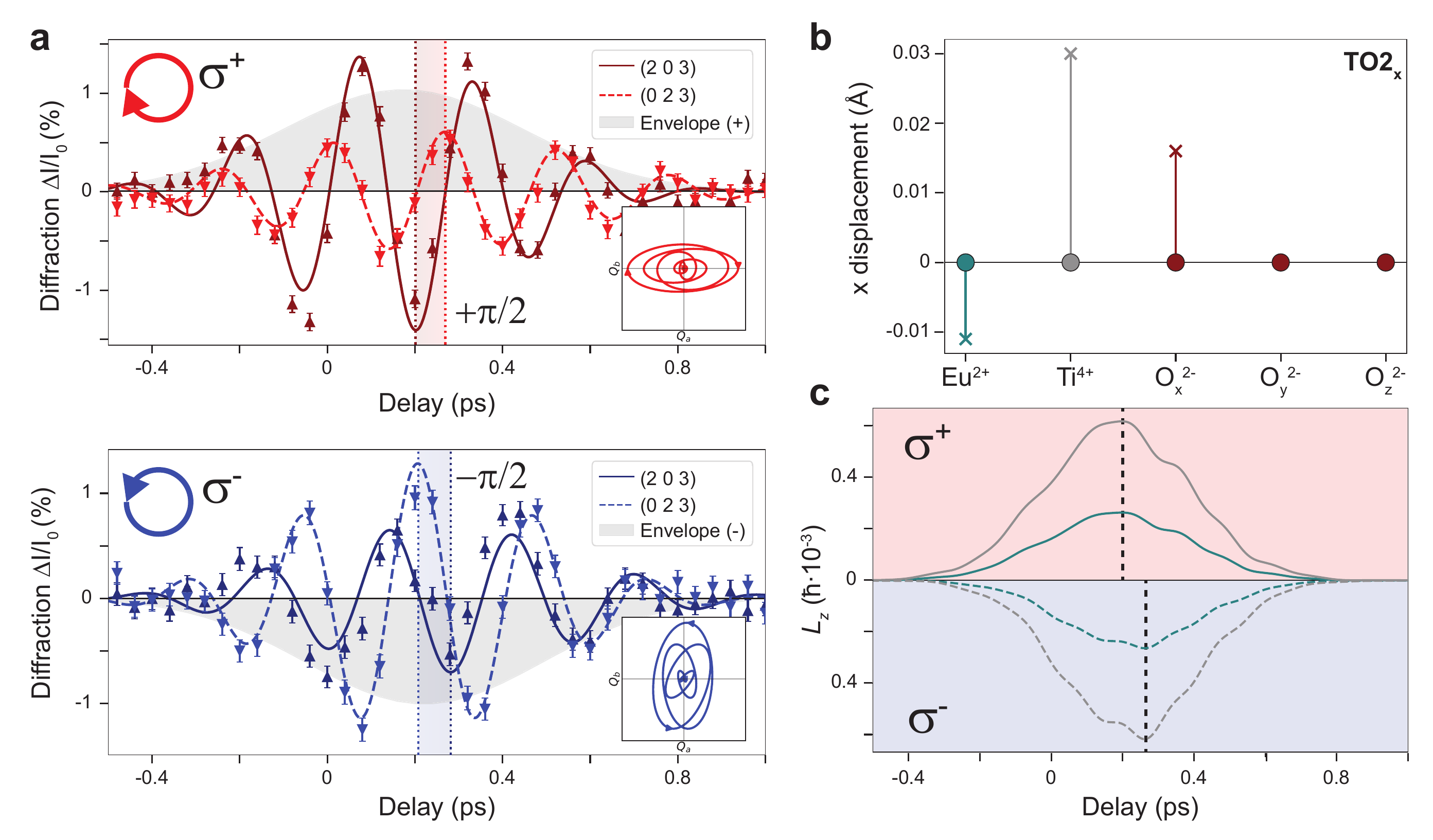}
\caption{\textbf{Observation of circular ionic motion from trXRD.}
(a) Change in the relative intensity of the $(2~0~3)$ and $(0~2~3)$ Bragg reflections during excitation with circular THz of opposite helicities. A clear $\pm\pi/2$ phase shift is seen for $\sigma^\pm$ THz excitation. Insets depict the projection of the $a$- and $b$-axis motion for each helicity of THz drive. Error bars represent standard error of the mean. (b) Depiction of the eigenvector from reference \cite{to22012phonon} for displacements along $x\parallel a$, with arbitrary real-space amplitude. The degenerate $y\parallel b$ component is obtained by the corresponding permutation of the Cartesian displacement components and oxygen sites ($\mathrm{O_x}\rightarrow\mathrm{O_y}$); see Supplemental. (c) Time-dependent angular momentum of each ion calculated from the diffraction data with fits and structure-factor calculations performed in the frozen phonon approximation. The upper half (shaded red) represents THz ($\sigma^+$) excitation and the lower half (shaded blue) represents THz ($\sigma^-$) with dashed vertical lines showing the maximal value of ionic angular momentum for both. Note the timing is shifted between the two circular drives due to the time shift of one of the constituent linear THz pulses used to create the circular polarizations.}
\end{figure}

At the XPP beamline of the LCLS \cite{lcls2015x}, we performed ultrafast hard X-ray measurements at the europium $L_2$-edge (7613 eV; see Fig. 3b). We used time-resolved XRD to measure the structural response of EuTiO$_3$ with circular THz excitation. By pumping with both helicities of THz excitation and measuring multiple diffraction peaks, we were able to directly quantify the circular ionic motion at room temperature. Specifically, we probed the $(2~0~3)$ and $(0~2~3)$ Bragg peaks (pseudocubic basis), monitoring the atomic displacements projected along the $a$- and $b$-axes, respectively, under identical experimental conditions (identical sample orientation; see Fig. 1d). In Fig. 2a, the relative intensity change of these diffraction peaks is plotted for positive and negative helicity of the circular pump; top and bottom, respectively. From these data, a clear $\pm\pi/2$ phase shift is observed between the orthogonal diffraction responses for the two helicities of THz excitation, which is a direct observation of circular (elliptical) ionic motion, as the in-plane angle between the reciprocal lattice vectors for the two reflections is also $\pi/2$. 

To provide a smooth time-varying function for further analysis, we performed phenomenological fits to the data with a carrier-envelope function (see Supplemental), which are shown as the solid and dashed lines in Fig. 2a. Using the THz excitation fields measured prior to the experiment and the corresponding calculated fields inside the film, we extract the excited phonon frequencies with a driven damped harmonic oscillator model (see Extended Data Fig. 3 and Supplemental). The fitted frequencies are approximately 3.8 and 4.4 THz along the $a$- and $b$-axes, respectively. In particular, the $b$-axis frequency closely matches the $\sim$ 4.5 THz thin-film TO2 resonance measured along the DSO $[1\bar{1}0]_\mathrm{o}$ direction, which is the same direction as our $b$-axis \cite{thinfilm2010strong}. Although spectral weight of the THz is also present near the lower-frequency TO1 mode, fits including both TO1 and TO2 find no appreciable TO1 contribution to the measured diffraction response (see Supplemental), supporting a predominantly TO2 assignment. The absence of a TO1 response could be due to the strong damping of the mode near the ferroelectric transition. Considering the excitation of the orthogonal TO2 components along the $a$- and $b$-axes, we can use the eigenmode(s) shown in Fig. 2b to perform structure-factor calculations to determine the coherent ionic motions during circular THz excitation (see Supplemental for more details). For these calculations, we approximate the mode eigenvectors using the bulk EuTiO$_3$ phonons from reference \cite{to22012phonon} since we are above the ferroelectric transition temperature (see Supplemental).

In the absence of any non-Maxwellian fields, the magnetic moment produced by dynamical multiferroicity will originate classically from the angular momentum of each ion. With the atomic motion derived from the trXRD and structure-factor calculations (see Supplemental), we can calculate the angular momentum of each ion as $L_i(t)=\mathbf{r}_i(t)\times\mathbf{p}_i(t)$, and sum the corresponding ionic magnetic moments to obtain the magnetic moment per unit cell:
\begin{equation}
\mu(t)=\sum_i\frac{q_i}{2m_i}L_i(t),\label{eqMoment}
\end{equation}
where $\mathbf{r}_i$ is the displacement of ion $i$ from its equilibrium position during the excitation, $\mathbf{p}_i$ is its instantaneous momentum, and $\frac{q_i}{2m_i}$ is the gyromagnetic ratio of each ion. The resulting time-dependent angular momentum of each ion is shown in Fig. 2c. At the time of maximum total angular momentum, we obtain a classical ionic magnetic moment of $\mathbf{3\times10^{-8}~\mu_B}$ per unit cell, averaged over the magnitudes for the two helicities of the drive. As this quantity is derived from the measured atomic motion alone and does not account for additional non-Maxwellian contributions, it provides the classical ionic contribution associated with the axial motion in EuTiO$_3$. We note that a recent publication has performed similar measurements of circular ionic motion when exciting the TO1 mode in SrTiO$_3$ \cite{mankowsky2026quantifying}.

\section{Europium magnetic response}\label{sec3}

\begin{figure}[!t]
\centering
\includegraphics[width=1\linewidth]{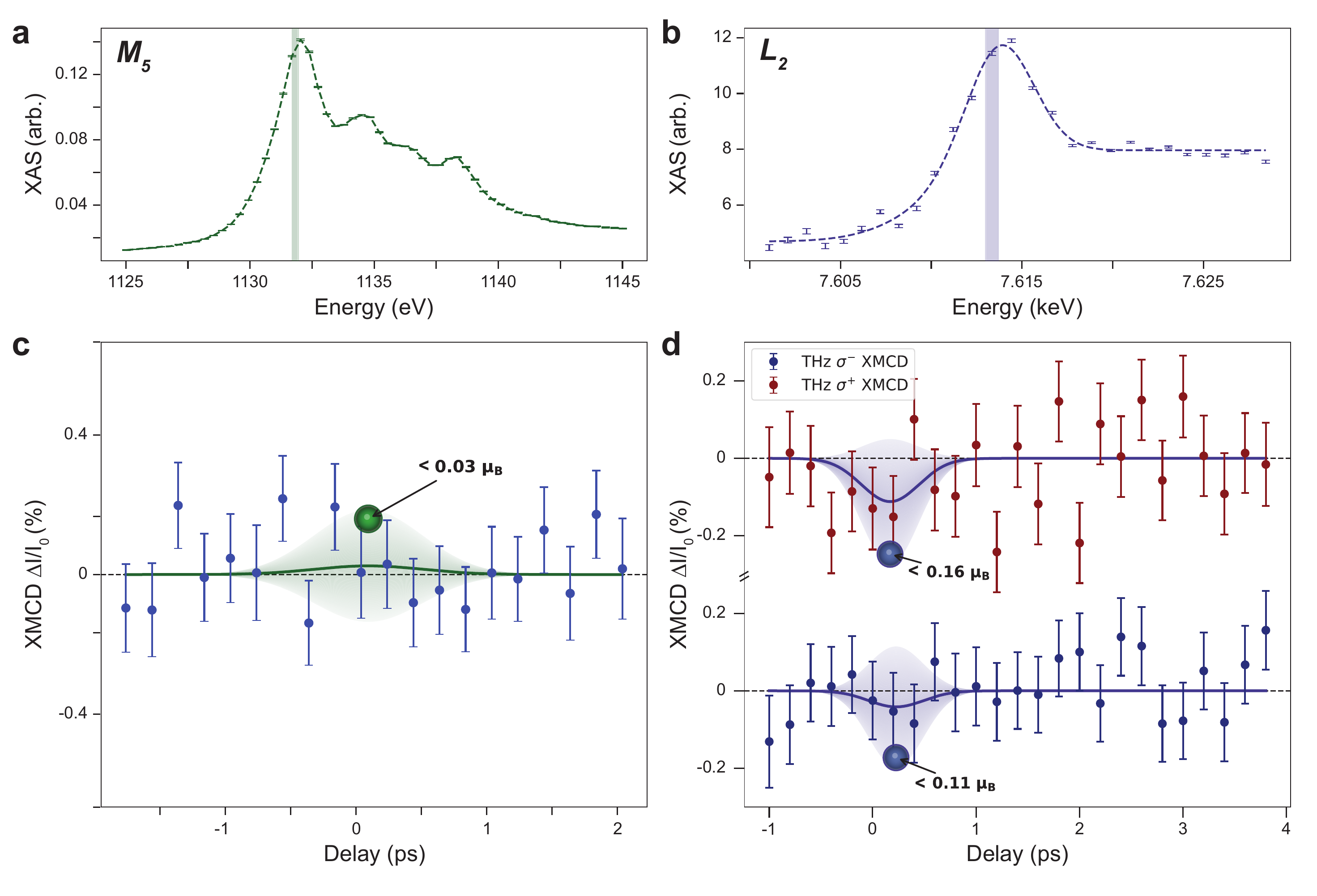}
\caption{\textbf{Element-resolved XMCD bounds on the induced Eu$^{2+}$ moment.}
(a) XAS spectrum of the europium $M_5$-edge taken at Furka in fluorescence mode. The shaded vertical bar represents experimental energy at approximately 1131.8 eV. Error bars represent the standard error. (b) XAS spectrum of the europium $L_2$-edge taken at XPP in fluorescence mode. The shaded vertical bar shows the chosen experimental energy for both XRD and XMCD around 7.613 keV. Error bars represent the standard error. (c),(d) Room-temperature XMCD response of the $M_5$-edge and $L_2$-edge during circular THz excitation with error bars representing standard error. Fits from a phonon model are overlaid for one THz helicity in the soft X-ray experiment and two THz helicities in the hard X-ray experiment. For all cases, no discernible signal is observed, evidenced by the two-sided 95\% confidence interval of the fit amplitude including zero. The spheres mark the peak responses corresponding to the one-sided 95\% confidence bound used to determine the upper limit on the induced moment.}
\end{figure}

The classical ionic moment determined from the circulating motion is only $3\times10^{-8}~\mu_B$ per unit cell. However, we can search for additional contributions to the magnetization from an alignment of Eu$^{2+}$ spins which could originate from strong effective non-Maxwellian fields. Thus, we turn to XMCD to look for any magnetic signatures of the Eu$^{2+}$ ion. The europium ion has a large paramagnetic moment that might become significantly spin-polarized in the non-equilibrium state. We performed XMCD measurements at both the europium $L_2$ edge at XPP in LCLS \cite{lcls2015x} and the $M_5$ edge at the Furka endstation in SwissFEL \cite{abela2019athos,swissfelFurka}. Measurements were performed at room temperature for each edge, as well as low-temperature measurements at $100~\mathrm{K}$ for the $L_2$-edge and $30~\mathrm{K}$ for the $M_5$-edge (see Extended Data Fig. 4 and Supplemental). Due to the dipole selection rule, transitions at these edges probe either the \emph{5d} ($L_2$) or \emph{4f} ($M_5$) electrons and associated magnetic moments of the europium ion, providing insight into both orbital and spin magnetization dynamics. X-ray absorption spectra at each edge taken in fluorescence mode are shown in Fig. 3a and 3b. To search for the ultrafast magnetization, we record the pump-induced change in fluorescence for alternating $c^+$ and $c^-$ X-ray helicities. The difference between the two helicities provides the time-resolved XMCD response and therefore a direct probe of an induced Eu magnetic moment.

The ultrafast change in Eu$^{2+}$ XMCD at room temperature is shown for both absorption edges in Fig. 3c and 3d with a time bin of 200 fs (see Methods). We note that no appreciable signal is seen from the data, and that the statistics are consistent with a null result (see Supplemental for more details). For the $L_2$-edge, we have direct access to the circular phonon envelope from the diffraction data that was simultaneously collected (see Fig. 2a). Using this envelope, we fit the XMCD amplitude and find that the measurement is consistent with no detectable change in the XMCD signal with $95\%$ confidence (see Fig. 3d). Specifically, the two-sided $95\%$ confidence interval of the fitted amplitude includes zero, while the corresponding one-sided $95\%$ confidence bound is used to determine the upper limit (see Supplemental). Using this upper bound, we find that the magnetic moment is at most $0.11~\mu_B$ when averaging the two THz helicities together and normalizing to the XMCD values found for Eu$^{2+}$ in the literature \cite{rogalev1997spin}. For the $M_5$ edge, we can similarly construct a circular phonon envelope to fit the data after simulating the phonon responses from the driven damped harmonic oscillator parameters derived from diffraction and the calculated internal THz fields. From these fits, we find the measurement is consistent with no measurable change in the XMCD with $95\%$ confidence (see Fig. 3c), and that the magnetic moment is at most $0.03~\mu_B$, where the moment is again calculated with respect to XMCD values found in the literature \cite{holroyd2004properties}. The exact details of these fitting procedures and statistical analysis can be found in the Supplemental.

\section{Response mechanisms and magnitudes}\label{sec4}

To better understand the implications of these results, we first consider the timescales of the optical response compared with the dynamics of the lattice. Our measurements show a helicity-dependent optical response and coherent circular ionic motion, while no corresponding Eu$^{2+}$ XMCD response is detected. In principle,  there are two distinct sources from which a magnetic signal (real or effective) can originate during a circular THz excitation \cite{DM2017dynamical,biggs2026ultrafast}. The first is the phonon-driven response described by dynamical multiferroicity (DM):
\begin{equation}
\mathrm{\mu_{DM}}(t)\propto\mathbf{Q}(t)\times\mathbf{\dot{Q}}(t),\label{eq1}
\end{equation}
where $\mathbf{Q}(t)=(Q_a(t),Q_b(t),0)$ is a vector containing the amplitudes of the orthogonal phonon components along $a$ and $b$, which create the circular ionic motion we have directly measured. Another possible process, which originates directly from the electronic response to the excitation field itself, is the inverse Faraday effect (IFE):
\begin{equation}
\mathrm{\mu_{IFE}}(t)\propto\mathbf{E}(t)\times\mathbf{\dot{E}}(t),\label{eq2}
\end{equation}
where $\mathbf{E}(t)=(E_x(t),E_y(t),0)$ is a vector describing the THz electric field inside the EuTiO$_3$ film with $x\parallel a$ and $y\parallel b$ (see Supplemental).

If we construct fits of the optical signal using both of these models, we find that the IFE model produces better agreement with our data than the DM model (see Fig. 4a). This demonstrates that the optical response is more consistent with the field-driven temporal envelope than the longer-lived phonon-driven envelope, implying that magnetic effects could occur on slightly faster timescales. The inverse Faraday effect is a well-studied phenomenon \cite{kimel2005ultrafast,van1965optically}, which is essentially based on an effective field originating from the transfer of orbital and spin angular momentum via spin-orbit coupling through a virtual state transition \cite{pershan1966theoretical,popova2011theory,popova2012theoretical}. This provides a microscopic pathway that could, in principle, induce a magnetic moment. To investigate the possible influence of the IFE mechanism on the europium spins, we re-bin our XMCD data to 40 fs for the $L_2$ edge and 66 fs for the $M_5$ edge, and fit the data with the envelope given by the internal THz excitation fields and Eq. \ref{eq2} (see Extended Data Fig. 4 and Supplemental). This model similarly finds that the measurement is consistent with no measurable change in the XMCD with $95\%$ confidence, and gives upper bounds of 0.04 and 0.13 $\mu_B$ for the $M_5$ and $L_2$ edges, respectively. In addition, we note that the $L_2$ and $M_5$ edges probe Eu \emph{5d} and \emph{4f} states, which might couple very differently to a non-Maxwellian field. The \emph{4f} states are well-localized pure spin moments, whereas the \emph{5d} states are hybridized with the moving oxygen ions and also carry non-zero orbital angular momentum, meaning our bounds are significant for both local (spin) and orbital moments. The bounds extracted from both edges nevertheless correspond to the total equivalent Eu magnetic moment when calibrated against static XMCD reference measurements.

\begin{figure}
\centering
\includegraphics[width=1\linewidth]{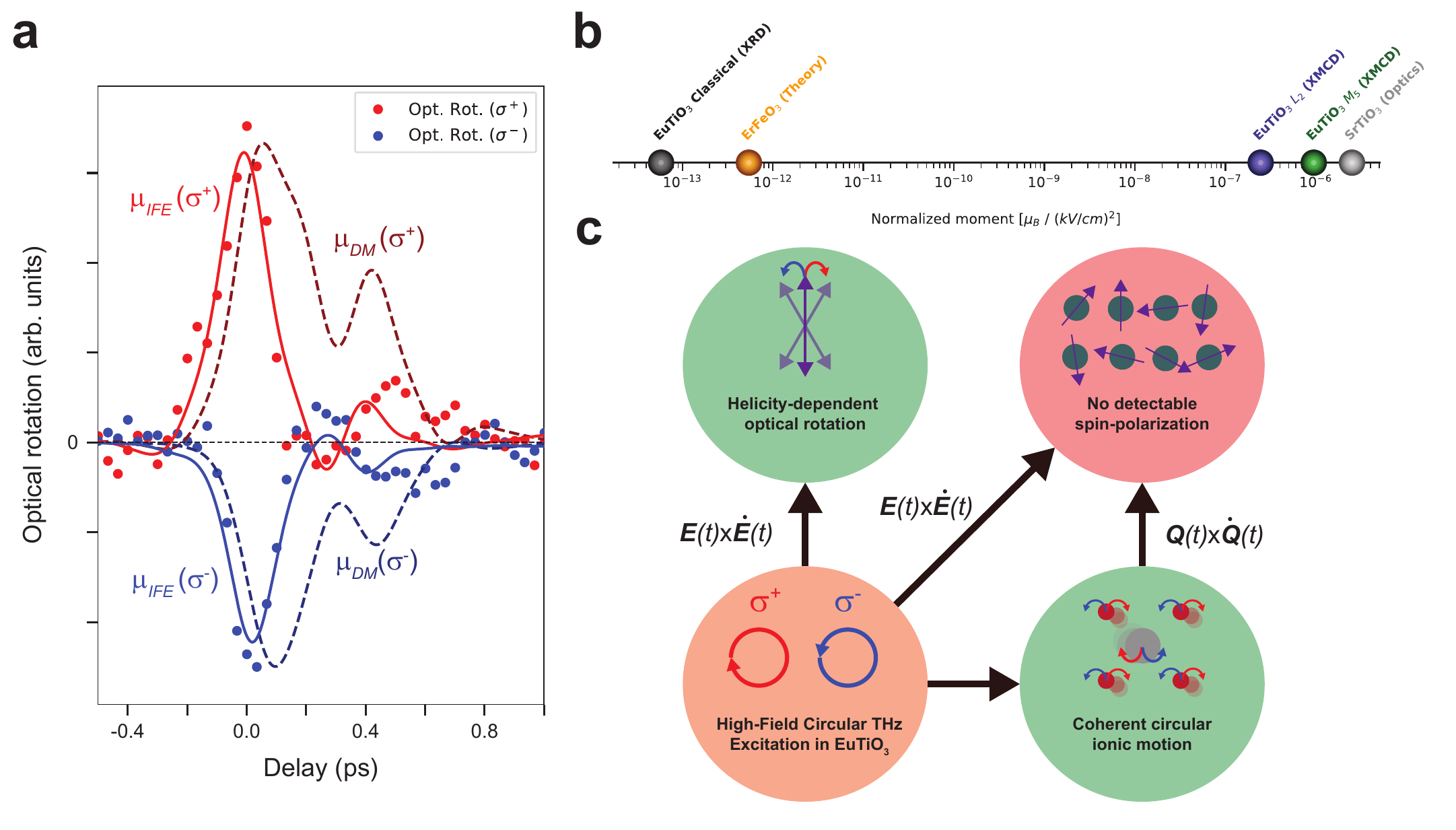}
\caption{\textbf{Comparison of mechanisms and moments.}
(a) Helicity-dependent optical rotation data overlaid with the IFE envelope calculated from the internal THz excitation fields using Eq. \ref{eq2}, and the DM envelope obtained by simulating the phonon response to the internal THz fields using the phonon parameters derived from the diffraction data and Eq. \ref{eq1}. The optical response is fit with the electric-field envelope with subscript $IFE$ and is seen to reproduce the data well when allowing the timing and amplitude of the envelope as fit parameters. The phonon envelope with subscript $DM$ is seen to be delayed with respect to the optical envelope and to have features which do not match the data well, including a broader response and no zero crossing. (b) Comparison of magnetic moments normalized by the square of their peak incident circulating electric field. The XMCD upper bounds remain many orders of magnitude above the classical ionic moment determined from the mechanical angular momentum in the diffraction experiment. In addition, we plot the reported moment in SrTiO$_3$ \cite{basini2024terahertz} induced by an approximately 200 kV/cm circulating peak electric field as well as the original theoretical prediction for ErFeO$_3$ with a simulated 10 MV/cm peak electric field \cite{DM2017dynamical}. (c) Distinguishing the causal chain of events when exciting EuTiO$_3$ with high-field circular THz. The driving field directly produces a rotation of the polarization state of an optical probe as well as coherent circular ionic motion. However, neither the driving electric field nor circular ionic motion creates a detectable spin polarization.}
\end{figure}

Since the structural response does not appear significant for the apparent Faraday rotation response observed with optics, we provide XMCD fits assuming an IFE temporal envelope for the low-temperature measurements (see Extended Data Fig. 4). The moment bounds here are nearly identical, but with the difference that the magnetic susceptibility of the paramagnetic Eu ion is significantly higher at lower temperature, indicating that the sensitivity to any real or effective magnetic fields acting on the Eu spin should be strongly enhanced. Given the relation $\mathbf{M}=\chi\mathbf{B}_{\mathrm{ext}}/\mu_0$, where the moment is directly proportional to the magnetization, we can calculate the external field which would produce our maximum possible moments. For the upper bound at the more sensitive $M_5$-edge, we find an equivalent external field of approximately $100~\mathrm{mT}$ at 30 K (see Supplemental for details). Here, we use the Curie-Weiss magnetic susceptibility that well describes the unquenched free spin moment of a Eu$^{2+}$ ion in its paramagnetic state with a magnetic transition to the ferromagnetic state at approximately 4 K \cite{laguta2017magnetic}.

When comparing the results of the optical measurements with those performed with X-rays, we must consider that the THz pumps have different peak fields and somewhat different waveforms (see Methods and Extended Data Fig. 2). From Eqs. \ref{eq1} and \ref{eq2}, we observe that both effects are proportional to the incident electric field squared in the linear-response regime. Therefore, we normalize the calculated magnetic moments by the square of the peak incident circulating THz electric field (see Supplemental for further details). In Fig. 4b we show a normalized magnetic moment for the classical ionic motion found from diffraction, as well as the normalized upper bounds of the XMCD measurements at the $M_5$ and $L_2$ edges. We compare these normalized values with bulk SrTiO$_3$, which has the large reported moment with an incident peak circular field of approximately 200 kV/cm \cite{basini2024terahertz}, as well as ErFeO$_3$ from the original theory work \cite{DM2017dynamical}. We further provide a temperature-dependent estimate of the equivalent external field which would produce such a moment for each case, normalized in the same way, in the Supplement. 

Since the epitaxially grown thin film has a different phonon spectrum compared to the bulk, a precise match with bulk SrTiO$_3$ results is not expected. Specifically, the excitation in SrTiO$_3$ is attributed to the soft mode (TO1), while we attribute our dynamics predominantly to the coherent motion of the TO2 mode. We note that a recent XRD work demonstrated that the TO1 mode of SrTiO$_3$ has its angular momentum dominated by the three oxygen atoms within the unit cell \cite{mankowsky2026quantifying}, whereas the TO2 mode which we assume to excite here does not have an angular momentum contribution from oxygen. For comparison, assuming excitation of the TO1 mode increases the calculated classical ionic moment by approximately a factor of 300, to order $10^{-5}~\mu_B$, which remains many orders of magnitude below our XMCD sensitivity (see Supplemental). We further note that a recent optical experiment provided a magnetic moment much closer to the original order of magnitude predicted by theory, likely because of the use of the material's magnetic susceptibility when converting their polarization rotation into a magnetic moment \cite{biggs2026ultrafast}. The reported moment for SrTiO$_3$ was obtained without the use of the magnetic susceptibility, which may contribute to the differences shown in Fig. 4b. In understanding the origin of the optical signals, we note that recent nonlinear phononics theory and experiments have shown that the excitation of geometrically chiral phonons can produce natural optical activity which would have an optical signal similar to dynamical multiferroicity \cite{zeng2025photo,zeng2025photo2,chen2025ultrafast}. While the chirality of axial $\Gamma$-point phonons is small (negligible $k$), it is possible that some optical activity would arise during the circular motion of the atoms.

This work establishes that an apparent magnetic optical rotation signal is not equivalent to a significant spin polarization when driving circular ionic motion with high-field THz pulses. The resulting causal chain is summarized in Fig. 4c, where the circular THz drive produces both a helicity-dependent optical response and coherent circular ionic motion, while neither is accompanied by a detectable Eu spin polarization. Furthermore, our results imply that such signals in optical measurements may arise from moments below our XMCD sensitivity or from origins other than magnetism. In general, considering the tiny classical contribution from ionic motion that breaks time-reversal symmetry, large-amplitude circular ionic motion driven by THz pulses is insufficient on its own to create significant transient magnetization without a distinctive coupling to the electronic system (e.g. crystal field excitations \cite{chaudhary2024giant,juraschek2022giant}), despite clear helicity-dependent signatures in the optical data.

\section{Methods}\label{secMethods}

\subsection{Sample growth}\label{subsecSample}

The EuTiO$_3$ thin film was epitaxially grown on a (110)-oriented DyScO$_3$ single-crystal substrate by pulsed laser deposition. Deposition was carried out at a substrate temperature of 720~$^\circ$C under a molecular oxygen partial pressure of $10^{-6}$ mbar. The growth was monitored in situ by reflection high-energy electron diffraction (RHEED), whose intensity oscillations confirmed a layer-by-layer growth mode and allowed the deposition to be controlled in real time, yielding a film with a thickness of approximately 33 nm. Following deposition, the film was annealed in situ for 45 min at the growth temperature to improve the film crystallinity.

\subsection{Production of circular high-field THz}\label{subsecTHz}

High-field circular THz pulses were generated with two distinct methods for the set of experiments in this work.

\subsubsection{Split and delay (optical and $L_2$-edge measurements)}

\noindent For both the $L_2$-edge and optical experiments, the split and delay method for generating high-field circular THz was used (see Extended Data Fig. 1). The signal and idler pulses coming out of an optical parametric amplifier (OPA) are used independently to create linear THz pulses through optical rectification in state-of-the-art organic crystals (PNPA: (E)-4-((4-nitrobenzylidene)amino)-N-phenylaniline) \cite{rader2022new}. The pulses are then recombined with the use of a THz polarization optic which transmits one linear polarization and reflects the other. A variable delay between the vertical and horizontal THz components allows us to control the ellipticity of the circular THz pump pulse. The variable delay was applied to the signal arm for the $L_2$-edge measurements and to the idler arm for the optical measurements (see Extended Data Fig. 1 and Supplemental). This high-field circular (elliptical) pulse is then sent through a series of parabolic mirrors to be refocused onto the sample position. X-ray and optical probes are incident through a hole in the final parabolic mirror to be at normal incidence. A global delay stage provides relative delays between the probe and THz pump, giving the dynamics of interest in this experiment.

\subsubsection{Quartz waveplate ($M_5$-edge measurements)}

\noindent For the $M_5$ edge measurements, high-field THz pulses were generated with the use of optical rectification of an OPA signal pulse in the organic crystal DSTMS (4-N,N-dimethylamino-4’-N’-methyl-stilbazolium 2,4,6-trimethylbenzenesulfonate) \cite{mutter2007linear}. These linearly polarized THz pulses were then sent through a zero-order quartz waveplate with a thickness of 550 $\mu$m, which is tuned to provide circular THz light at roughly 2.8 THz. This THz light is then sent through a series of parabolic mirrors before being focused onto the sample with a spot size of $\sim$400 $\mu$m. The optical laser had 100 fs pulse duration and a 100 Hz repetition rate. 

\subsection{X-ray methods}\label{subsecX}
For all experiments, the THz pump and X-ray/optical probe pulses are at near-normal incidence of the sample so that the c-axis of the pseudocubic basis is parallel to the optical axis. This ensures the THz can excite orthogonal phonon modes in plane and ensures the XMCD/Faraday effect is sensitive to a change in the out-of-plane magnetization (see Fig. 1). For both X-ray edges, data were processed shot-to-shot with the use of a timing-tool and incident X-ray intensity normalization. The data were filtered to select for shots with desirable parameters, and empirical error bars for a single global delay stage scan (pump relative to probe timing) were generated by bootstrapping the shots within a specific time bin. These statistics were in turn used to generate the standard error when combining multiple scans (see Supplemental for more information). For the diffraction data, regions of interest (ROI) on the pixel detector were defined and integrated to provide the one-dimensional $\Delta I/I_0$ data.

\subsubsection{$L_2$-edge}

\noindent At the XPP end-station, X-ray pulses were monochromatized using a diamond (111) monochromator and tuned near the Eu $L_2$ edge. This configuration provided an energy bandwidth of 0.6–0.7 eV and a pulse duration of approximately 30–40 fs. The incident photon energy was calibrated using the Eu $L_2$ X-ray absorption spectrum measured from the sample in fluorescence mode. The X-ray beam was focused to a spot size of 100 $\mu$m on the sample. X-ray diffraction and fluorescence were recorded simultaneously using a Jungfrau1M and ePix100 detector, respectively. Circular X-ray polarization was generated by sending an initially linearly polarized X-ray pulse through a 960 $\mu$m-thick diamond phase retarder operated near the (220) Laue reflection. The helicity was switched by detuning the crystal to opposite angular offsets with a motorized stage.

\subsubsection{$M_5$-edge}

\noindent At the Furka end-station, X-ray absorption spectra were measured in fluorescence mode using a photodiode detector. Soft X-ray probe pulses were supplied by the Athos beamline of SwissFEL and tuned to the Eu $M_5$ edge, near 1130 eV, and fine-tuned using the fluorescence signal from the sample. The beam was generated in self-amplified spontaneous emission mode and monochromatized to a bandwidth of approximately 0.5 eV. The X-ray pulse duration was about 40 fs, and the experiment was operated at a 100 Hz repetition rate. The X-ray helicity was alternated between circular $(c^+)$ and $(c^-)$ polarization using the Apple X undulators and was focused to a spot size of approximately 200 $\mu$m on the sample. Transmission gratings provided shot-to-shot diagnostics of the incident X-ray intensity and the pump-probe arrival time was found by spectral encoding with a SiN$_x$ membrane. These measurements were used for normalization and timing correction without perturbing the time-resolved XAS (XMCD) measurement.

\clearpage
\bibliography{main_bib_v2}

\section*{Extended Data}

\begin{figure}[H]
\centering
\includegraphics[width=1\linewidth]{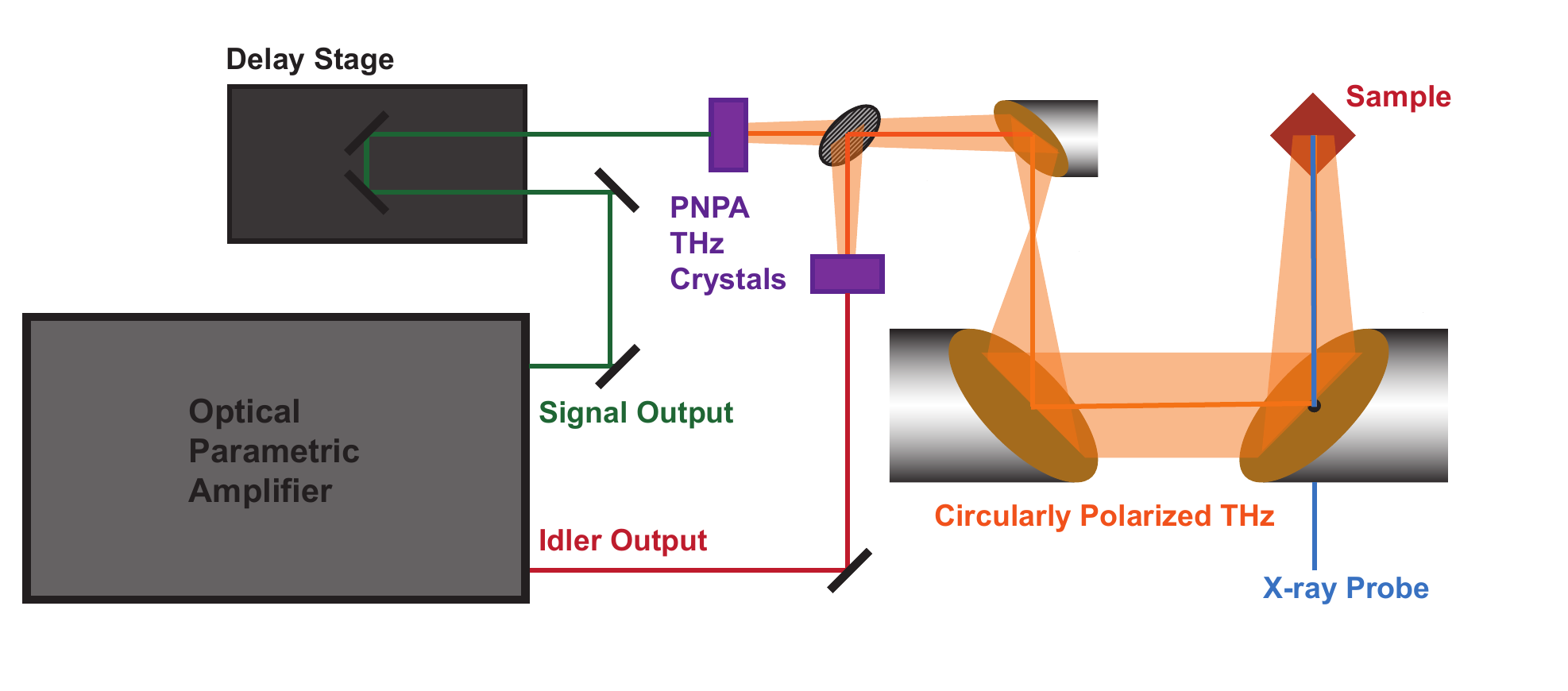}
\extendedfigcaption
{Split-and-delay generation of circular THz pulses}
{Signal and idler outputs from an optical parametric amplifier are used to generate THz pulses in an organic crystal. The pulses have orthogonal linear polarizations and are recombined with a variable delay to create a predominantly circular THz pump pulse. Here, the signal arm is illustrated with the variable delay as in the LCLS experiment, while the Supplement shows the corresponding layout for the BYU optical experiments.}
\label{fig:extended_split_delay}
\end{figure}

\begin{figure}
\centering
\includegraphics[width=.9\linewidth]{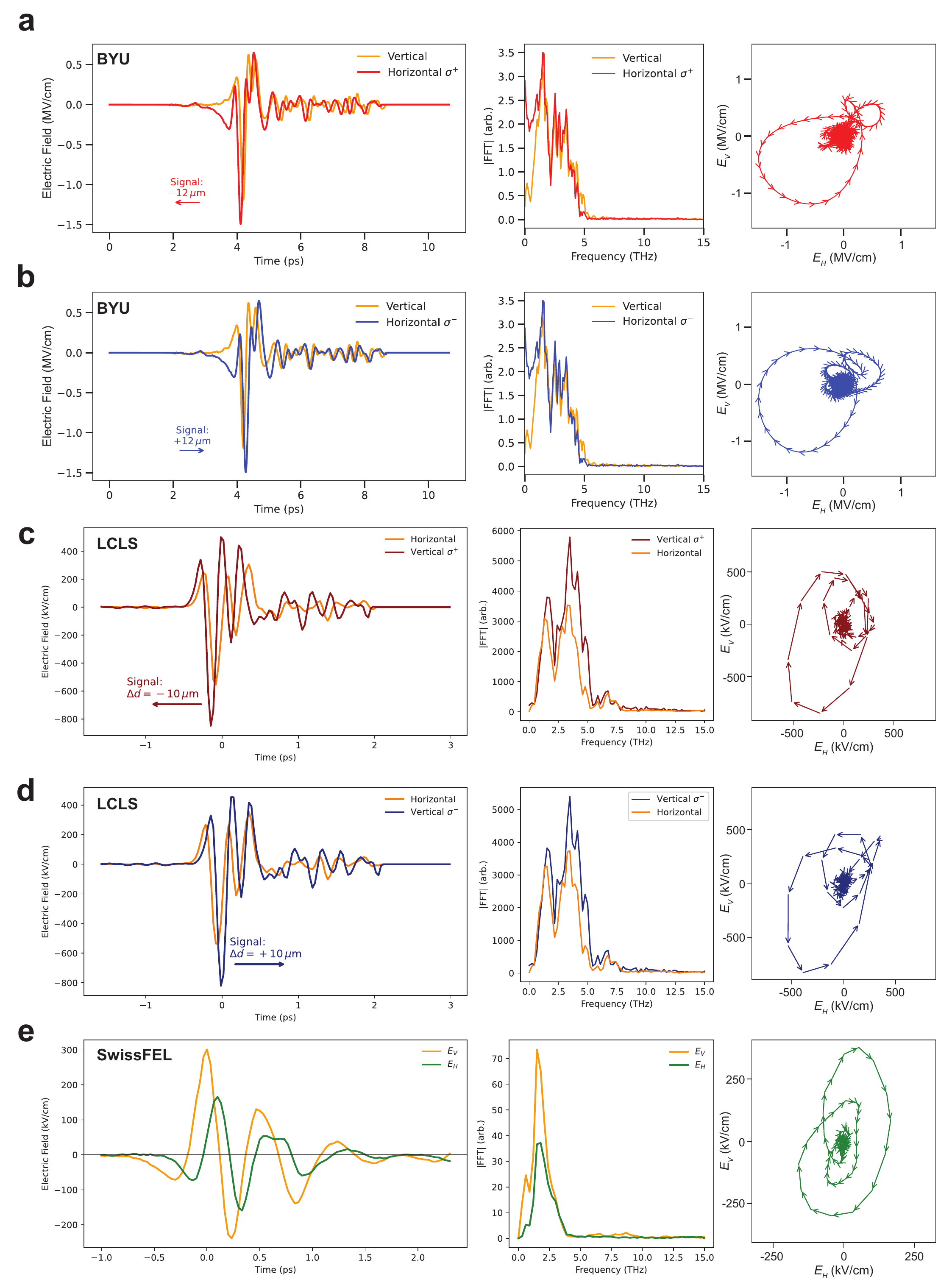}
\extendedfigcaption
{Experimental THz fields}
{(a),(b) High-field circular THz electric fields used for the optical rotation experiment at the Brigham Young University (BYU) laboratory. Note that the horizontal polarization is variably delayed here. (c),(d) High-field circular THz electric fields used for the $L_2$ edge experiment at the XPP endstation at LCLS. Note that the vertical polarization timing is variable here meaning the $\sigma^\pm$ definitions are reversed compared with the BYU fields. (e) THz electric field used at the Furka endstation at SwissFEL. This circular THz is produced with a quartz waveplate and a different organic crystal compared with BYU and LCLS.}
\label{fig:extended_thz}
\end{figure}

\begin{figure}
\centering
\includegraphics[width=.9\linewidth]{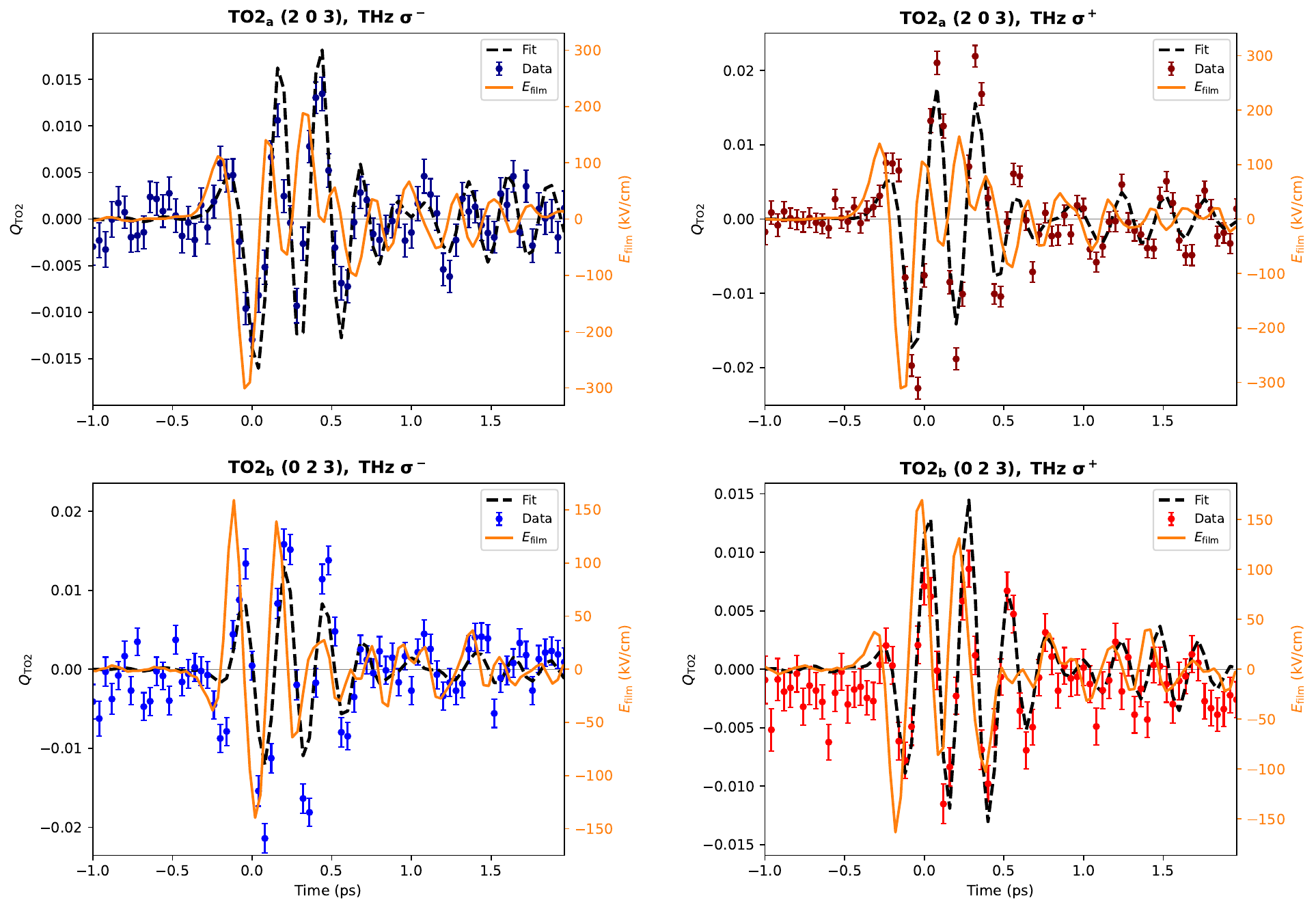}
\extendedfigcaption
{Damped harmonic oscillator global fitting}
{Using the internal THz excitation fields, we can perform global fits to the diffraction responses within a driven damped harmonic oscillator model. This yields
$f_a = 3.8\,\mathrm{THz}$ and
$\gamma_a = 2.2\,\mathrm{ps}^{-1}$ for the $a$-axis mode, and
$f_b = 4.4\,\mathrm{THz}$ and
$\gamma_b = 3.5\,\mathrm{ps}^{-1}$ for the $b$-axis mode. See Supplemental for more details.}
\label{fig:extended_DHO}
\end{figure}

\begin{figure}
\centering
\includegraphics[width=1\linewidth]{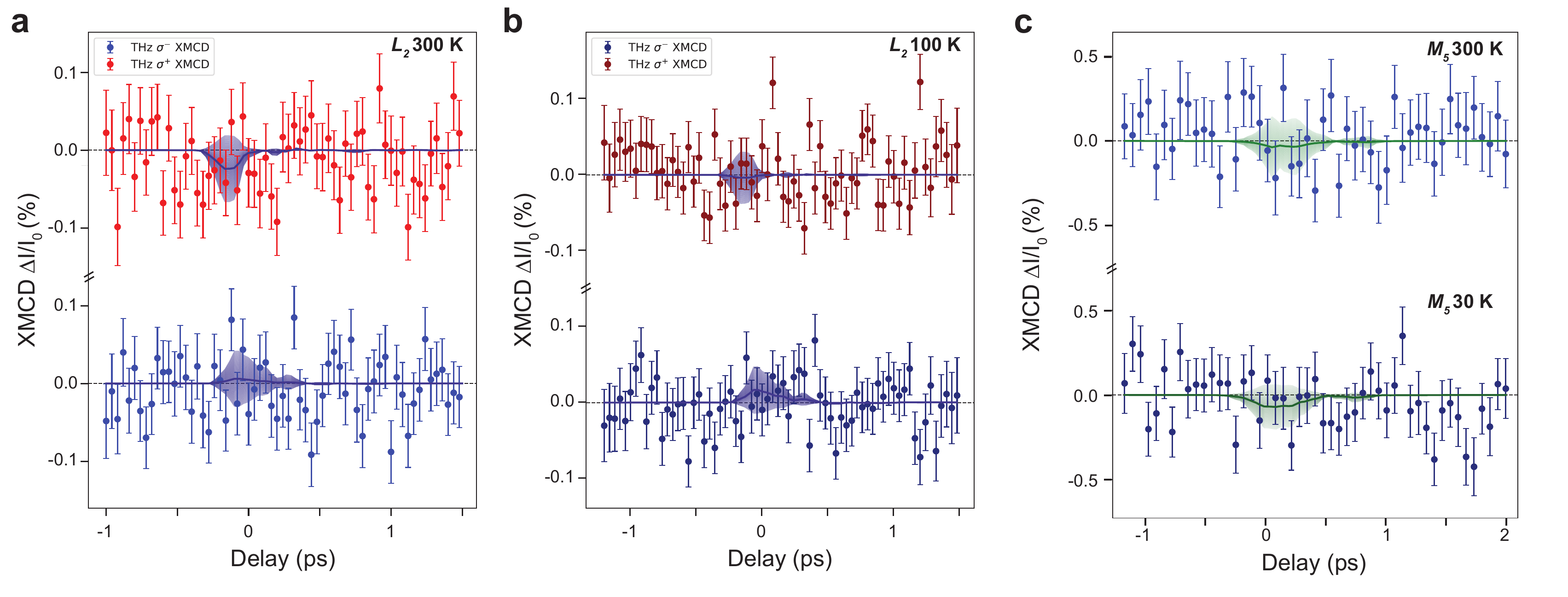}
\extendedfigcaption
{IFE envelope fits of XMCD}
{(a),(b) $L_2$ edge XMCD measurements with IFE envelope fits given by Eq. \ref{eq2} at room temperature and 100 K. (c) $M_5$ edge XMCD measurements with IFE envelope fits given by Eq. \ref{eq2} at room temperature and 30 K.}
\label{fig:extended_IFE}
\end{figure}

\end{document}